\documentclass[reprint,groupedaddress,amsmath,amssymb,aps,pra,floatfix]{revtex4-1}
\pdfoutput=1
\usepackage{amsmath,amssymb,graphicx,mathrsfs,amsfonts,braket,blkarray}

\newcommand{\bs}[1]{\boldsymbol{#1}}
\DeclareMathOperator{\tr}{\text{Tr}}

\begin{document}

\title{Laser Cooling by Sawtooth Wave Adiabatic Passage} 

\author{John P. Bartolotta}
\affiliation{JILA, NIST, and University of Colorado, 440 UCB, 
Boulder, CO  80309, USA}
\author{Matthew A. Norcia}
\affiliation{JILA, NIST, and University of Colorado, 440 UCB, 
Boulder, CO  80309, USA}
\author{Julia R. K. Cline}
\affiliation{JILA, NIST, and University of Colorado, 440 UCB, 
Boulder, CO  80309, USA}
\author{James K. Thompson}
\affiliation{JILA, NIST, and University of Colorado, 440 UCB, 
Boulder, CO  80309, USA}
\author{Murray J. Holland}
\affiliation{JILA, NIST, and University of Colorado, 440 UCB, 
Boulder, CO  80309, USA}

\date{\today}

\begin{abstract}
\noindent We provide a theoretical analysis for a recently demonstrated cooling method. Two-level particles undergo successive adiabatic transfers upon interaction with counter-propagating laser beams that are repeatedly swept over the transition frequency. We show that particles with narrow linewidth transitions can be cooled to near the recoil limit. This cooling mechanism has a reduced reliance on spontaneous emission compared to Doppler cooling, and hence shows promise for application to systems lacking closed cycling transitions, such as molecules. 
\end{abstract}


\maketitle 


\section{INTRODUCTION} \label{intro}

\noindent The use of coherent light to cool the motion of particles such as atoms to sub-Kelvin temperatures has greatly expanded the capabilities of atomic and molecular physics \cite{metcalf}.  The earliest and simplest mechanism demonstrated and understood was Doppler cooling \cite{dopp1,dopp2}, which relies on preferential absorption between counterpropagating lasers, followed by spontaneous emission.  While Doppler cooling is simple and robust, the large number of spontaneous emissions involved has several drawbacks: spontaneously emitted photons impart random momentum kicks to the particle that cause diffusion and limit the achievable temperatures. These spontaneous emission events can also result in the particle falling into internal states that are no longer near resonance with the applied light so that the cooling ceases. In the case of atoms, the latter can be mitigated by adding a small number of additional lasers to ``repump'' into cooled states, but in the case of molecules the number of uncooled states may be so large that this approach becomes a significant challenge \cite{molecular_cooling}. Doppler cooling can reach the recoil temperature (set by the recoil energy from a single emitted photon) for narrow linewidth transitions such that $\gamma < \omega_r$, where $\gamma$ and $\omega_r$ are the optical transition linewidth and recoil frequency respectively.  However, the cooling time scale in this parameter regime is long, scaling inversely with $\gamma$.

Several approaches to laser cooling have been developed to mitigate the negative effects of spontaneous emission \cite{adams}. In sub-Doppler cooling mechanisms such as Sisyphus cooling \cite{sisyphus}, the energy removed per spontaneously emitted photon is large, allowing lower temperatures to be reached. In Raman sideband cooling \cite{sideband}, tight confinement of the atom can suppress the effect of momentum recoil associated with spontaneous emission, enabling cooling to nearly the ground state of the external potential well. Cavity-cooling techniques \cite{cavity_cooling} can be used to reduce free-space spontaneous emission by causing preferential decay to a desired state via the output coupler of an optical resonator.

Recently, we have experimentally demonstrated a new cooling mechanism, named ``SWAP cooling'' (sawtooth wave adiabatic passage cooling) in which particles are coherently driven between ground and excited states of a narrow-linewidth optical transition by counterpropagating, frequency-swept lasers.  Spontaneous emission is still critical in order to remove entropy from the system to achieve steady-state cooling, but by coherently driving a particle between its ground and excited state multiple times, large amounts of energy can be removed per spontaneous emission.  This provides a way to generate significant forces and to reach low temperatures while maintaining a large velocity capture range. The approach is largely insensitive to perturbations such as laser frequency drifts, magnetic fields, and AC Stark shifts.  These factors have made SWAP cooling a useful experimental technique for cooling atomic strontium using its 7.5~kHz linewidth, dipole-forbidden ${^1S}_0 \rightarrow {^3P}_1$ transition \cite{exp}. A related procedure has been used in the generation of a new form of magneto-optical trap \cite{swap_mot}. Sub-Doppler cooling of $^{87}$Rb was recently observed using SWAP cooling with two-photon Raman transitions between ground hyperfine states \cite{greve}. In general, the approach is applicable to any atomic species, especially alkaline-earth-like atoms that possess intercombination transitions. Further, the reduced reliance on spontaneous emission may make SWAP cooling a useful tool for cooling molecules that have narrow linewidth optical transitions, such as the $160$ kHz linewidth ${{X}^{2}}\Sigma \rightarrow A{{^{\prime} }^{2}}{{\Delta }_{3/2}}$ transition in YO \cite{collopy}.

Here, we present a detailed theoretical analysis of SWAP cooling.  We explore the minimum achievable temperature as well as the various laser-particle interactions that affect particle dynamics. We investigate its capture range and the forces involved in the cooling process. We show that for appropriate parameters, SWAP cooling can be used to cool to near the recoil limit.  We also simulate the rate at which spontaneous photons are emitted during cooling, confirming that the amount of energy removed per spontaneous emission event can greatly exceed the limits of Doppler cooling.


\section{basic mechanism} \label{qualitative}

\begin{figure}[!htb]
  \begin{center}
  \includegraphics[width=\linewidth]{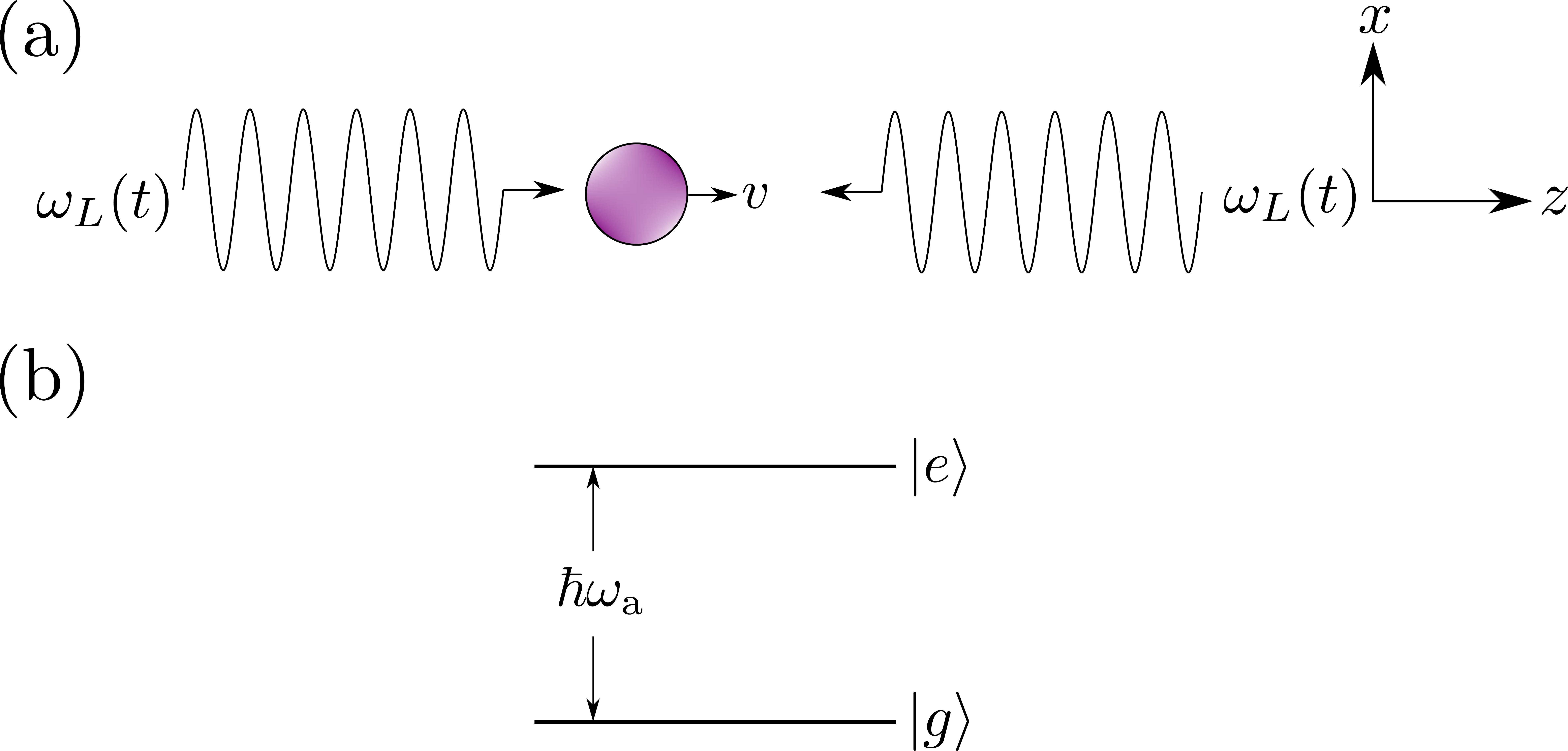}
  \end{center}
\caption{(a) Spatial setup of the standing wave (two counterpropagating lasers with frequency $\omega_L(t)$) and particle (circle) with velocity $v$ in the lab frame. (b) Energy level diagram of the bare internal states, separated by energy $\Delta E = \hbar \omega_\text{a}$.}
  \label{fig:setup}
\end{figure}

\noindent The main mechanism for momentum removal in SWAP cooling is the coherent transfer of a particle toward zero momentum via adiabatic passage. Momentum is removed by time-ordered stimulated absorption and emission of photons caused by interaction with a standing wave formed by counter-propagating laser beams. To describe the cooling, we consider a two-level particle with internal states $\ket{e}$ and $\ket{g}$, separated in energy by~$\hbar \omega_\text{a}$, and one dimension of motional freedom along the $\hat{z}$-direction (see Figure \ref{fig:setup}). We choose to represent the external degree of freedom in the momentum basis and with label $\ket{p}$, where $p$ denotes a momentum eigenstate such that $\hat{p} \ket{p} = p \ket{p}$. 

As shown in Figure 2, the laser frequencies, $\omega_L(t)$, follow an asymmetric sawtooth waveform with full range~$\Delta_s$ and period $T_s$ so that the slope of the sawtooth ramp is $\alpha \equiv \Delta_s / T_s$. The sawtooth frequency ramp is centered on the transition frequency, $\omega_\text{a}$, and the frequency is  linearly ramped from below to above $\omega_\text{a}$. We approximate the wavenumber, $k$, to be fixed throughout the sweeping sequence. This also implies that the recoil frequency, $\omega_r = \hbar k^2 / 2 m$, is fixed, where $m$ is the mass.

\begin{figure}[!htb]
  \begin{center}
  \includegraphics[width=\linewidth]{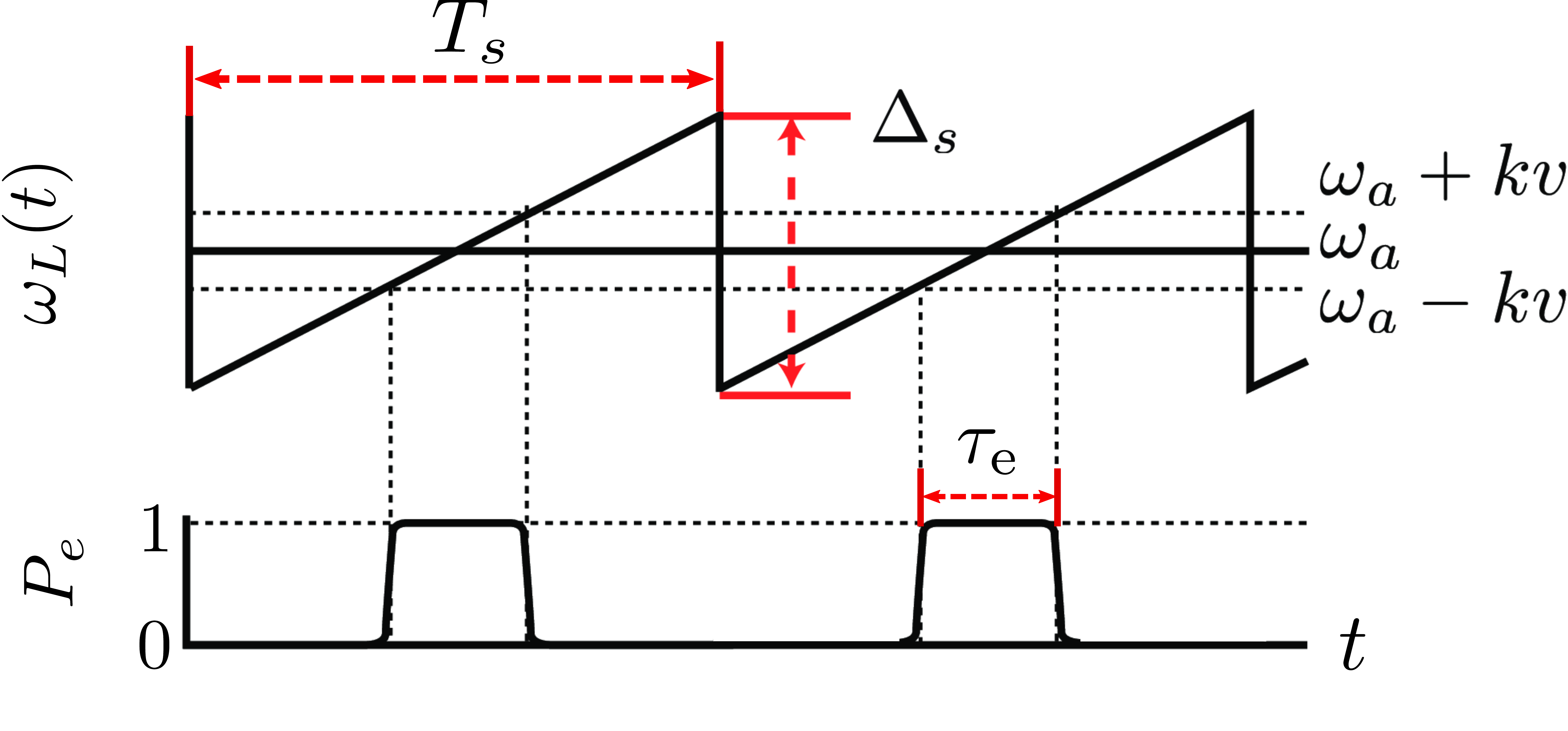}
  \end{center}
\caption{Top: The laser frequency, $\omega_L(t)$, as a function of time. The approximate resonance frequencies for a particle with velocity $v$ are labeled. The sweep range is chosen to be large enough such that both beams will become resonant with the particle at some time during the sweep. Bottom: The ideal excited-state fraction, $P_e$, in the adiabatic regime. The particle remains in the excited state for a time interval $\tau_\text{e}$.}
  \label{fig:sweep}
\end{figure}

The direction of particle motion matters since Doppler shifts set the time-ordering of which beam first interacts with the particle.  To understand this time-ordering, consider a particle initially in the state $\ket{g,p_i}$.  The Doppler shift ensures that the counter-propagating beam is the first to sweep across the transition frequency. If we first focus on the case with $p_i > 0$, this adiabatically transfers the particle to the state $\ket{e , p_i - \hbar k}$ via stimulated absorption. After some time, the co-propagating beam will achieve resonance, adiabatically transferring the particle to $\ket{g, p_i-2 \hbar k}$ via stimulated emission. For $p_i < 0$, the mapping is $\ket{g,p_i} \rightarrow \ket{e, p_i+\hbar k} \rightarrow \ket{g,p_i + 2 \hbar k}$. In either case, the particle is transferred closer to zero momentum. The net result is the removal 2$\hbar k$ of momentum and transfer back to the particle's initial internal state without spontaneous emission. Subsequent sweeps would then continue to remove momentum in units of 2$\hbar k$. Figure \ref{fig:sweep} shows this process by illustrating the ideal excited state fraction, $P_e$, over two sweeps. 

The particle's internal state at the beginning of a sweep is a crucial factor in determining the impulse it receives from each laser. If the particle instead begins a sweep in the excited state, it is transferred away from zero momentum. Hence, one of the roles of spontaneous emission is to ensure that the particle begins a sweep in the ground state. This is achieved by requiring an imbalance between the times spent inside and outside of the two resonances, which corresponds to
\begin{equation}
\label{eq:captureRange}
 \Delta_s > 4 |kv|.
\end{equation}
This condition also enforces the requirement that the sweep range, $\Delta_s$, is large enough for both laser beams to achieve resonance with the particle. 

There are several other conditions that must be satisfied in order to realize SWAP cooling. In order to ensure a low probability of decay during the time interval the particle is in the excited state, $\tau_\text{e}$, we require the condition
\begin{equation}
\label{eq:keep_excited}
   \tau_\text{e} \ll \frac{1}{\gamma}.
\end{equation}
Importantly, the Rabi frequency of each laser beam, $\Omega_0$, and the laser frequency sweep rate, $\alpha$, must satisfy the condition
\begin{equation}
\label{eq:atrans}
	\kappa \equiv \frac{\Omega_0^2}{\alpha} \geq 1,
\end{equation}

\noindent so that there is a substantial probability, $P_a$, for an adiabatic transition at each resonance \cite{zener}:
\begin{equation}
    P_a = 
        1- \exp \left[- \frac{\pi}{2} \frac{\Omega_0^2}{\alpha}\right].
\end{equation}

\noindent We shall refer to $\kappa$ as the adiabaticity parameter. Any~$\kappa$ that satisfies Eq.~(\ref{eq:atrans}) is said to be within the adiabatic regime, and any $\kappa$ that does not satisfy Eq.~(\ref{eq:atrans}) is said to be within the diabatic regime.

\section{SYSTEM DYNAMICS} \label{dynamics}

\noindent The quantum master equation 

\begin{equation} \label{eq:ME}
    \frac{d \hat{\rho}}{dt} =
    \frac{1}{i \hbar} [\hat{H},\hat{\rho}] + \hat{\mathcal{L}}(\hat{\rho})
\end{equation}
\noindent governs the time evolution of the density matrix, $\hat{\rho}$, whose Hilbert space includes both the particle's internal and external degrees of freedom. The Hamiltonian, $\hat{H}$, captures the coherent dynamics, while the Lindblad superoperator,  $\hat{\mathcal{L}}(\hat{\rho})$, captures the incoherent dynamics due to spontaneous emission and the associated recoil.

The operators $\hat{z}$ and $\hat{p}$ describe the particle's position and momentum, respectively. The excited and ground states, $\ket{e}$ and $\ket{g}$, form a pseudo-spin 1/2 system.  The usual raising ($\hat{\sigma}^{+}= \ket{e}\bra{g}$) and lowering ($\hat{\sigma}^{-}=\ket{g}\bra{e}$) operators, along with the usual Pauli spin operators, $\hat{\sigma}^{x,y,z}$, whose coordinate labels are implicitly understood to refer to the pseudo-spin space ({\em e.g.}, ~$\hat{\sigma}^{z}= \ket{e}\bra{e}-\ket{g}\bra{g}$, {\em etc.}), operate on the internal states.

The electric field of the applied lasers is polarized along the quantization axis, $\hat{x}$, (see Figure \ref{fig:setup}) and is described in first quantization by the operator
\begin{align}
\label{eq:electricfield}
    \hat{\bs{E}}(\hat{z},t) = 
    \hat{x} E_0
    \Bigl[
        \cos 
        \left(
            k \hat{z} + \eta(t)
        \right) 
        + \cos
        \left(
            k \hat{z} - \eta(t)
        \right)
    \Bigr].
\end{align}
Here, the amplitude of each standing wave is $E_0$, and the time-dependent accumulated phase of the laser field,~$\eta(t)$, from initial time $t_0$ is 
\begin{equation}
    \eta(t) \equiv \int_{t_0}^t \omega_L(t') \, dt'\,.
\end{equation}

In the interaction picture defined by the free Hamiltonian
\begin{equation}
    \hat{H}_0(t) = 
        \frac{\hat{p}^2}{2m}
        + \frac{\hbar}{2} \omega_\text{a} \hat{\sigma}^z,
\end{equation}

\noindent the particle's non-dissipative dynamics, under the dipole and rotating wave approximations, is described by the interaction Hamiltonian
\begin{align} \label{eq:Hamiltonian}
    \hat{H} =
   \frac{\hbar}{2} \Omega_s  \cos
        \left(
            k\hat{z} + \frac{k\hat{p}}{m} t
        \right)
        \left(
    		\hat{\sigma}^+ e^{- i \theta(t)} + \text{h.c.} 
        \right),
\end{align}
where 
\begin{equation}
	\theta(t) \equiv
    \int_{t_0}^t \omega_L(t') - \omega_\text{a} \, dt'
    =\eta(t) - \omega_\text{a}t
\end{equation}

\noindent is the time-dependent phase of the laser field's detuning from resonance.  The standing wave's peak Rabi frequency, 
\begin{align}
    \Omega_s \equiv 
    	2\Omega_0 =  
        - \frac{2 \braket{e | \hat{\bs{d}} | g}  \cdot \hat{x} E_0}{\hbar},
\end{align}
characterizes the interaction strength of the electric field with the particle's electric dipole operator $\hat{\bs{d}}$.

The Lindblad operator,
\begin{align}
  \hat{\mathcal{L}}(\hat{\rho}) & =
  - \frac{\gamma}{2}
    \Bigl(
    \hat{\sigma}^+ \hat{\sigma}^- \hat{\rho}
    + \hat{\rho} \hat{\sigma}^+ \hat{\sigma}^- 
    - 2
    	\Bigl\{ 
        \frac{3}{5} \hat{\sigma}^- \hat{\rho} \hat{\sigma}^+ \notag \\
    & + \frac{1}{5}e^{i k \hat{z}} \hat{\sigma}^- \hat{\rho} \hat{\sigma}^+ e^{- i k \hat{z}}
    	+ \frac{1}{5}e^{- i k \hat{z}} \hat{\sigma}^- \hat{\rho} \hat{\sigma}^+ e^{ i k \hat{z}}
    	\Bigr\}
    \Bigr),
\end{align}
describes the effect of spontaneous emission. In order to keep the momentum distribution on a discretized grid, we have approximated the dipole radiation pattern to produce recoil of magnitudes $-\hbar k,0$, and $\hbar k$ along $\hat{z}$ with probabilities $\tfrac{1}{5}:\tfrac{3}{5}:\tfrac{1}{5}$, respectively \cite{castin}. We consider numerical and theoretical results of Eq.~(\ref{eq:ME}) in the following sections.


\section{dynamics in the high-velocity regime} \label{singlephoton}

\noindent The core mechanism whereby SWAP cooling removes momentum and energy from a particle's motion is most easily understood in a regime in which one can consider that the particle interacts sequentially with one traveling wave and then the other. We shall henceforth refer to this as the ``high-velocity regime." To define it, we must consider the time it takes to adiabatically transfer a particle with initial velocity $v_i$ between its internal states, which we call $\tau_\text{jump}$, as well as the time interval separating the two resonances, which we denote by $\tau_\text{res}$. In the adiabatic regime, it can be shown that $\tau_\text{res} = 2(kv_i-2\omega_r)/\alpha$ and $\tau_\text{jump} = 2 \Omega_0 / \alpha$ (see Appendix \ref{tres} and \cite{vitanov}, respectively). 

\begin{figure}[!htb]
  \includegraphics[width=\linewidth]{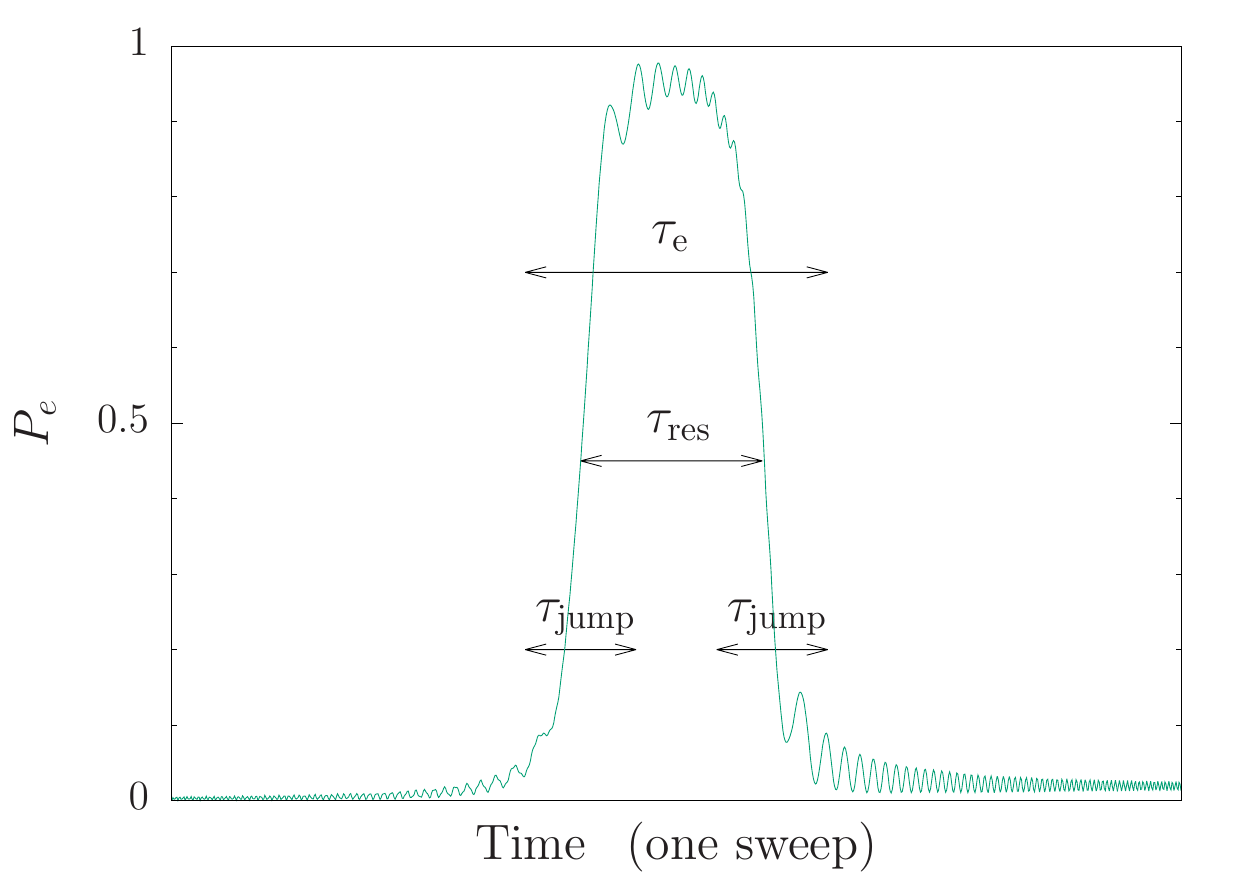}
  \caption{The excited state fraction, $P_e$, of a particle prepared in the state $\ket{g,10 \hbar k}$ over one sweep. Values, in units of $\omega_r$, are: $\Delta_s = 200, T_s = 22,$ and $\Omega_0 = 5$.}
  \label{fig:jumpstimjump}
\end{figure}

Figure \ref{fig:jumpstimjump} shows the time-ordering of these processes, as well as a measure of the total excited time, $\tau_\text{e}$, defined as the sum of $\tau_\text{jump}$ and $\tau_\text{res}$ in this regime.

\noindent Roughly half of each $\tau_\text{jump}$ overlaps with $\tau_\text{res}$. Therefore, to keep the resonances separated, we define the high-velocity regime to be the range of velocities that satisfy $\tau_\text{jump} < \tau_\text{res}$, or
\begin{equation}
\label{eq:highvelocityregime}
  |\Omega_0|< |kv_i - 2 \omega_r|. \qquad  (\text{high-velocity regime})
\end{equation}
It is important to emphasize that particles outside of the high-velocity regime may still be cooled under the SWAP cooling procedure. However, their dynamics are more difficult to describe and analyze qualitatively.

\subsection{Dressed state picture}
\label{dressed_state}

\noindent A convenient and intuitive way to understand adiabatic transfer is the dressed state formalism. Working in the laser frame, we diagonalize the Hamiltonian
\begin{equation} \label{eq:HamLaserFrame}
    \hat{H}(t) = 
        \frac{\hat{p}^2}{2m} 
        - \frac{\hbar}{2} \delta(t) \hat{\sigma}^z 
        + \frac{\hbar}{2} \Omega_s \cos (k \hat{z}) \hat{\sigma}^x,
\end{equation}
 at each instant in time. Here, $\delta(t) \equiv \alpha t$ is the laser detuning from resonance, since the detuning is linearly ramped from $-\Delta_s/2$ to $\Delta_s/2$. We track the evolution of the minimal set of eigenstates necessary to demonstrate the evolution of a particle that begins a sweep in the state $\ket{g,p}$ in the high-velocity regime. This set maps to the bare eigenstates
\begin{equation} \label{eq:four_states}
    \{ \ket{g,p}, \ket{e,p - \hbar k}, \ket{g, p-2 \hbar k}, \ket{e,p-3 \hbar k} \}
\end{equation}

\noindent in the limit of large detuning ($|\delta(t)| \gg | kv|$). The avoided crossing of a multi-photon process known as a Doppleron resonance, which does not affect the dynamics of a particle in the high-velocity regime, is also present (see Appendix \ref{doppleron} for details).

\begin{figure}[!htb]
  \begin{center}
  \includegraphics[width=\linewidth]{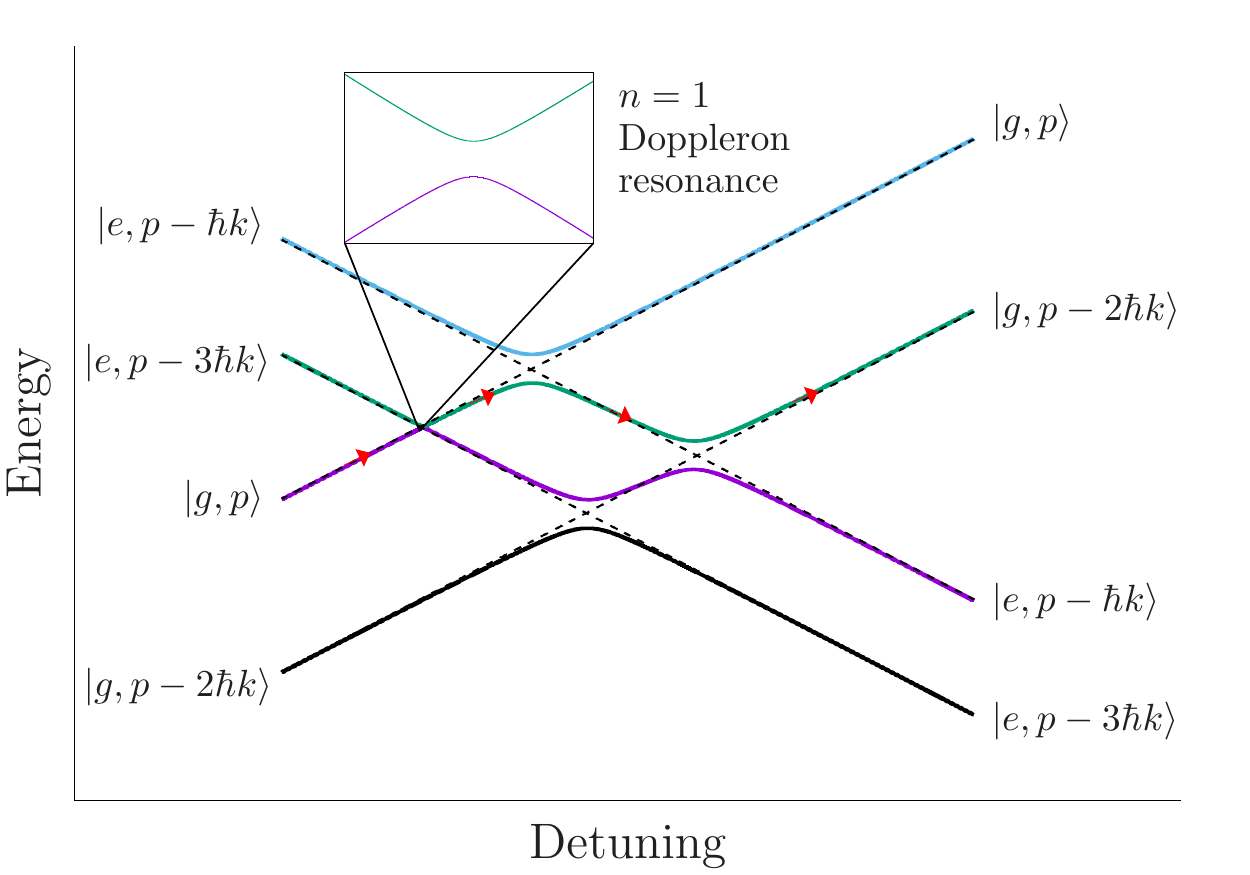}
  \end{center}
\caption{Eigenvalue energy versus detuning for the four coupled states given in Eq.~(\ref{eq:four_states}). The inset shows the splitting of an $n=1$ Doppleron resonance, which is small in the high-velocity regime. The red arrows identify the physical path being considered. Dashed lines show the evolution of the uncoupled states. Values, in units of $\omega_r$, are: $\Omega_0 = 2, \alpha = \omega_r = 1, T_s = 50$. $p = 4 \hbar k.$}
  \label{fig:4eval}
\end{figure}

Figure \ref{fig:4eval} shows the instantaneous eigenvalues as the detuning is linearly ramped. Starting in the state $\ket{g,p}$, the particle diabatically crosses the higher-order Doppleron resonance, thus being transferred into a different eigenstate. It then undergoes two adiabatic crossings, which correspond to the two resonances previously discussed, and ends up in the state $\ket{g,p-2 \hbar k}$, signifying the removal of $2 \hbar k$ of momenta. It is important to note that Figure \ref{fig:4eval} only depicts the correct evolution of a particle that follows the red arrows; in reality, the surrounding states would couple to states of higher and lower momenta. 

\subsection{Coherent dynamics}
\label{ideal_coherent}

\noindent In order to further illustrate the dynamics strictly due to coherent evolution, we numerically calculated the time evolution of the root-mean-square (rms) momentum of the particle, $p_\text{rms}=\sqrt{\langle \hat{p}^2\rangle}$, as it underwent SWAP cooling starting from the initial state $\ket{\psi_i} = \ket{g,10 \hbar k}$ and without spontaneous emission ($\gamma=0$). The Rabi frequency was chosen such that the simulation operated in the high-velocity regime for all momentum states $|p_i| > \hbar k$.

\begin{figure}[!htb]
  \centering
  \includegraphics[width=\linewidth]{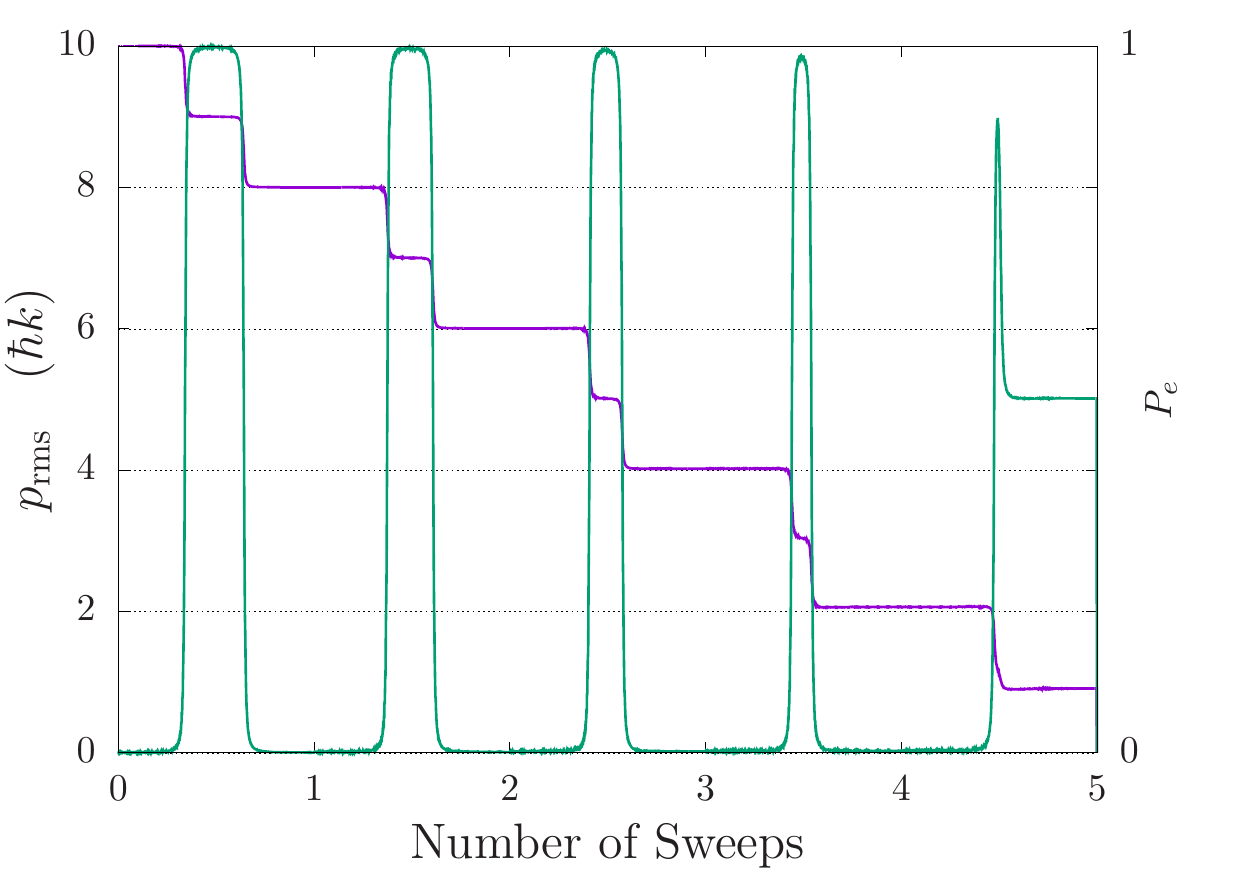}
  \caption{Root-mean-square momentum, $p_\text{rms} $ (step-like curve; magenta online), for a particle starting in the state $\ket{g,10 \hbar k}$ over five sweeps. The curve exhibiting rising and falling pulses shows the excited state fraction, $P_e$ (cyan online). The particle experiences a significant reduction in $p_\text{rms}$ and is left in a superposition of the internal states. Values, in units of $\omega_r$, are: $\Delta_s = 120, T_s = 1000,$ and $\Omega_0 = 1$.}
  \label{fig:coherenttransfer}
\end{figure}

Figure~\ref{fig:coherenttransfer} shows the rms momentum, $p_\text{rms}$, versus time, expressed in terms of the number of sweeps. The probability of finding the particle in the excited state, $P_e = |\langle e|\psi \rangle|^2$, is also shown on the right hand axis.  One sees that during each sweep the particle was adiabatically transferred to the excited state and then back to the ground state.  Each transition was accompanied by a reduction in the particle's momentum by $\hbar k$ for a total of $2 \hbar k$ per sweep.  As the momentum of  the particle was reduced, the time between transitions became shorter ({\em i.e.}, the width of the pulses became smaller). This is what one would expect since the velocity of the particle and therefore the accompanying Doppler shift is decreased.

By symmetry, and as confirmed by calculation, the rms momentum decreased in an identical manner for the state with opposite initial momentum, $\ket{g,-10 \hbar k}$.  Course graining over the individual sweeps, the effective force exerted on the particle drives it toward zero velocity, independent of its initial direction of motion. As a result, one should draw an important distinction between the force exhibited here and a ``slowing force'' that can be understood as applying a uniform translation to the momentum of all particles. Such slowing forces are often implemented in the context of slowing a molecular beam with chirped-frequency light \cite{frequency_chirp} or using rapidly varying electrostatic potentials as is done in Stark decelerators \cite{stark_decelerator}, but do not lead to steady-state cooling.

On the final sweep shown in Figure \ref{fig:coherenttransfer}, the particle approached zero momentum where the dynamics are modified. At the end of the final sweep, the particle had a 50$\%$ probability of occupying the $\ket{g,0}$ state, and a 25$\%$ probability of occupying each of the $\ket{e,\hbar k}$ and $\ket{e,-\hbar k}$ states, resulting in a final rms momentum of $p_\text{rms} = \hbar k/\sqrt{2}$ and a final excited state probability of $P_e =1/2$.  We will see that being left in the excited state at the end of the sweep is an important consideration for understanding the final equilibrium temperature.

\subsection{High-velocity regime dynamics including dissipation}

\noindent We have shown that under the influence of purely coherent dynamics, a particle prepared in its internal ground state can be sufficiently transferred to low momentum. In a realistic system, however, the presence of spontaneous emission restricts the amount of time a particle may remain in the excited state, Eq.~(\ref{eq:keep_excited}), which increases for higher initial momentum. Nevertheless, in the context of the high-velocity regime in the adiabatic limit, it can be shown that there is no upper bound on the momentum states that can be transferred to lower momentum via the SWAP cooling procedure, contingent that arbitrarily high Rabi frequencies and sweep rates are accessible. However, there exists a fundamental lower bound:
\begin{equation}
\label{eq:jmin_gamma}
	\left|\frac{p}{\hbar k} \right| \gtrapprox 
    	 1 + \frac{2 \kappa \gamma}{\omega_r}.
\end{equation} 

\noindent This motivates the use of SWAP cooling on a transition for which $\kappa \gamma/\omega_r$ is small. It is a requirement of adiabatic transfer to have $\kappa$ at least on the order of unity, so the experimentalist only has the freedom to vary the ramp slope $\alpha$ and Rabi frequency $\Omega_0$ accordingly. Regardless of experimental laser parameters, narrow linewidth transitions (on the scale of the recoil frequency) are preferable if the goal is to maximally cool the system.

\section{forces and capture range}
\label{forces}

\noindent For analyzing the cooling dynamics, it is useful to investigate the equivalent classical force exerted on a particle as a function of its velocity or momentum. We have chosen to describe this relationship by defining various quantities that provide information about the impulse imparted to the particle over a single sweep.

\subsection{Conservative Forces}
\label{conservative_forces}

\noindent One way to describe impulsive momentum kicks applied to the particle is the change in its rms momentum due to a single sweep:
\begin{equation}
\label{eq:rms_force}
	\Delta p_\text{rms} \equiv
    \sqrt{\bra{\psi_f} \hat{p}^2 \ket{\psi_f}}
      - \sqrt{\bra{\psi_i} \hat{p}^2 \ket{\psi_i}},
\end{equation}
where $\ket{\psi_i}$ is the state of the particle prior to the sweep and $\ket{\psi_f}$ is the state of the particle after the sweep. We describe the impulse in this way (rather than the average momentum) because the system exhibits Bragg oscillations, which are transitions between resonantly coupled momentum states $\ket{p} \leftrightarrow \ket{-p}$ (see Appendix \ref{bragg}). These oscillations yield an additional momentum change that is qualitatively different to the adiabatic transfer dynamics that we are interested in and does not contribute to the cooling process. The rms momentum is a measure that by construction excludes the effect of such Bragg oscillations.

\begin{figure}[!htb]
  \centering
  \includegraphics[width=\linewidth]{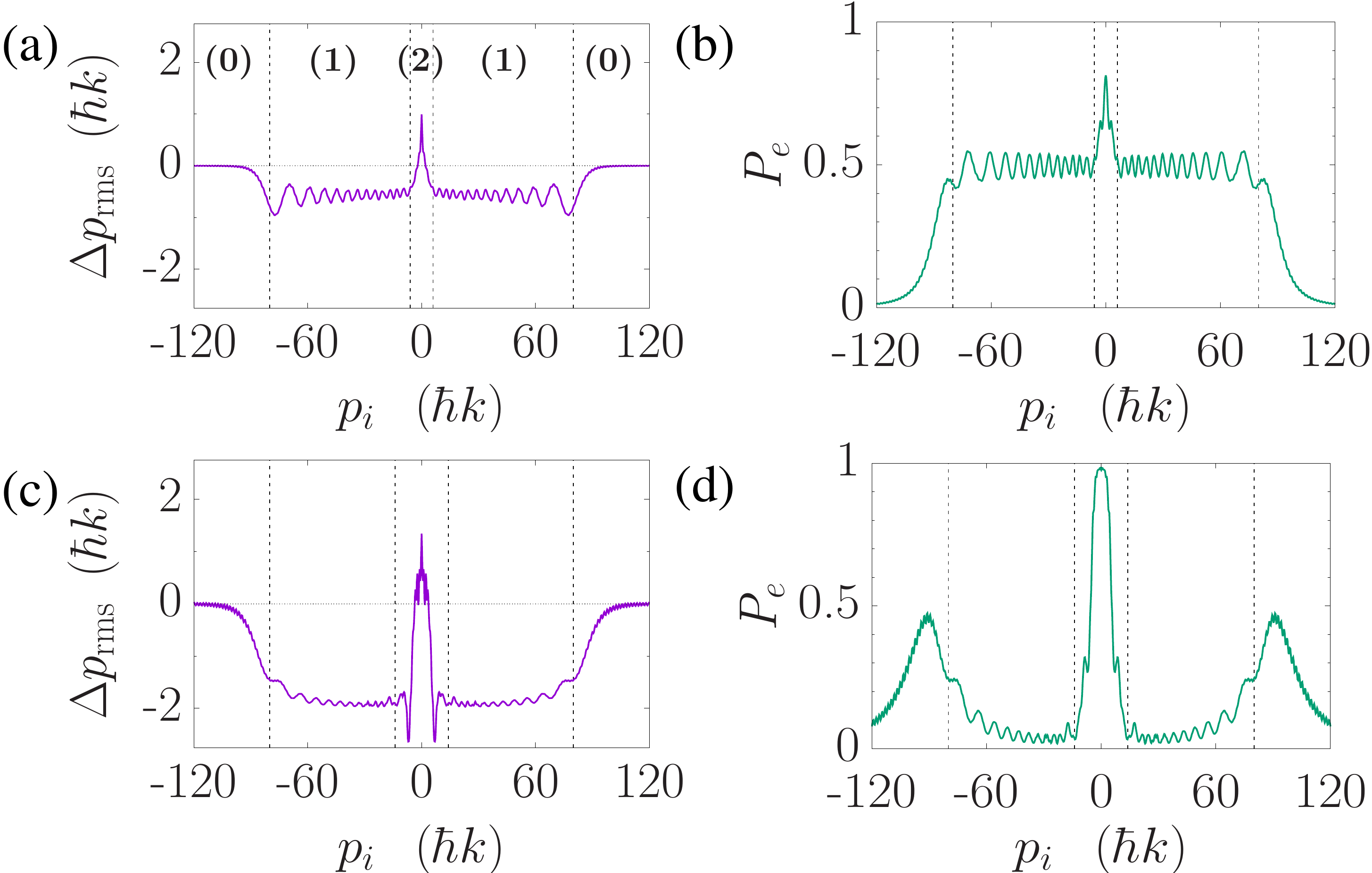}
  \caption{$\Delta p_\text{rms}$ vs. $p_i$ ((a) and (c)) and $P_e$ vs. $p_i$ ((b) and (d)) for a particle that started in the state $\ket{g,p_i}$ at the beginning of a sweep. The various momentum regions in Eqs.~(\ref{eq:regions}) are labeled in Figure \ref{fig:g_rms}(a) and are presented in all subplots by vertical, dashed lines. The horizontal, dashed line corresponds to $\Delta p_\text{rms}=0$. The adiabaticity parameter $\kappa$ lies in the diabatic regime for plots (a) and (b) ($\kappa = 0.5, \Omega_0 = 9.5 \omega_r$) and in the adiabatic regime for plots (c) and (d) ($\kappa = 4, \Omega_0=26.8 \omega_r$). Values common to all plots, in units of $\omega_r$, are: $\Delta_s = 360$, $T_s = 2$.}
  \label{fig:g_rms}
\end{figure}

\begin{figure}[!htb]
  \centering
  \includegraphics[width=\linewidth]{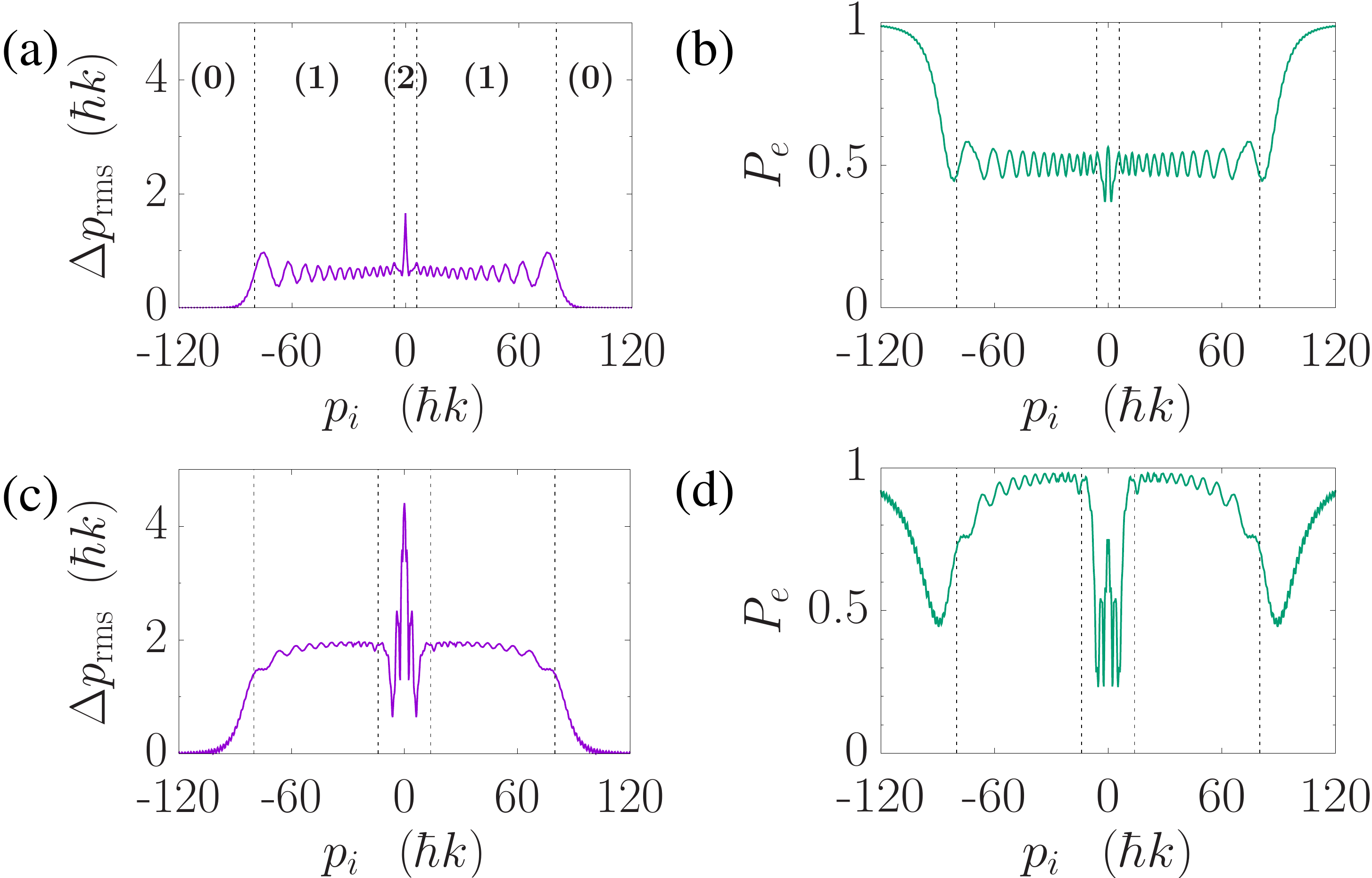}
  \caption{$\Delta p_\text{rms}$ vs. $p_i$ ((a) and (c)) and $P_e$ vs. $p_i$ ((b) and (d)) for a particle that started in state $\ket{e,p_i}$ at the beginning of a sweep. The various momentum regions in Eqs.~(\ref{eq:regions}) are labeled in Figure \ref{fig:e_rms}(a) and are presented in all subplots by vertical, dashed lines. The adiabaticity parameter $\kappa$ lies in the diabatic regime for plots (a) and (b) ($\kappa = 0.5, \Omega_0=9.5 \omega_r$) and in the adiabatic regime for plots (c) and (d) ($\kappa = 4, \Omega_0 = 26.8 \omega_r$). Values common to all plots, in units of $\omega_r$, are: $\Delta_s = 360$, $T_s = 2$.}
  \label{fig:e_rms}
\end{figure}

Figures \ref{fig:g_rms} and \ref{fig:e_rms} show the computed rms impulse $\Delta p_\text{rms}$ versus the initial momentum $p_i$ for a particle initially prepared in $\ket{\psi_i}=\ket{g, p_i}$ and $\ket{\psi_i}=\ket{e, p_i}$, respectively. The values chosen for the adiabaticity parameter, $\kappa = 0.5$ (top rows) and $\kappa=4$ (bottom rows), demonstrate the system's behavior in the diabatic and adiabatic regimes, respectively, and were varied by changing only the Rabi frequency, $\Omega_0$. The probability of being left in the excited state at the end of the sweep, $P_e$, is also provided. 

To aid in the description of the dynamics, we have labeled specific regions of momentum space with the symbols (0), (1), and (2), where ($i$) labels the maximum number of lasers the particle substantially interacts with at any time during the sweep. More specifically, a particle with initial velocity $v_i$ lies within the region defined by:
\begin{align}
	|k v_i| > \frac{\Delta_s}{2} & \qquad \text{region (0)} \notag \\
\label{eq:regions}
    |\Omega_0| < |k v_i| < \frac{\Delta_s}{2} & \qquad \text{region (1)} \\
    |k v_i| < |\Omega_0| & \qquad \text{region (2).} \notag
\end{align}

\noindent Note that region (1) roughly corresponds to the high-velocity regime, Eq~(\ref{eq:highvelocityregime}). Uninterestingly, particles in region (0) do not significantly interact with the laser field, so we restrict our discussion to regions (1) and (2).

For most states in region (1), the resulting $\Delta p_\text{rms}$ and $P_e$ after the sweep are roughly constant. The general results of a diabatic sweep in region (1), as seen in both Figures, may be interpreted as giving a low imparted impulse and a failure to return the particle to its initial internal state. In contrast, the general results of an adiabatic sweep in region (1) are an impulse of $\left|\Delta p_\text{rms}\right| \approx 2\hbar k$ and significant return to the initial internal state; these are the ideal coherent dynamics of the high-velocity regime as previously discussed. The highest momentum states within region (1) do not quite undergo this ideal behavior, even in the adiabatic regime, because the particle does not begin the sweep in an eigenstate of the Hamiltonian, Eq.~(\ref{eq:Hamiltonian}). 

As previously mentioned, Dopplerons and the ambiguous time-ordering of the two laser interactions can significantly modify the force in region (2) such that the physics is more complex. The effect of this complex behavior is clearly visible in Figures \ref{fig:g_rms} and \ref{fig:e_rms} as $|p_i|$ approaches zero. In particular, a particle within region (2) initially in the internal ground state will be transferred to the excited state after a sweep, which would then send it on a trajectory toward increasingly higher momentum. This motivates the requirement described by Eq.~(\ref{eq:captureRange}), which will on average reset the particle to the ground state for the next sweep via spontaneous emission.

\subsection{Forces including dissipation}
\label{dissipative_forces}

\noindent While Section \ref{conservative_forces} provides insight into the conservative forces in SWAP cooling, it does not include the dissipative features that ultimately lead to phase-space compression and equilibration. Moreover, we enforced specific state preparation at the beginning of each sweep. In order to explore the forces one would expect in the laboratory, {\em i.e.}, with $\gamma \neq 0$ and no specific state preparation, we define an average impulse as

\begin{equation}
\label{eq:avg_force}
	\Delta p_\text{avg} \equiv
    \tr \left[ \hat{p} \hat{\rho}_f\right]
    - \tr \left[ \hat{p} \hat{\rho}_i \right]
\end{equation}
with the constraint that the internal state populations are the same at the beginning and end of the sweep. The quantities $\hat{\rho}_i$ and $\hat{\rho}_f$ are the density operators associated with the initial and final particle states, respectively. We shall call these internal state populations the ``steady-state'' populations, $P_e^\text{ss}$ for each $p_i$. Note that steady-state here refers only to the internal state populations being equal before and after the sweep cycle; the momentum in general will change. This choice of representing the impulse is motivated by the desire to compare to other cooling methods where the internal populations reach a stationary situation, and to thereby allow investigation of the relationship between force and particle velocity in a more general context.

\begin{figure}[!htb]
  \centering
  \includegraphics[width=\linewidth]{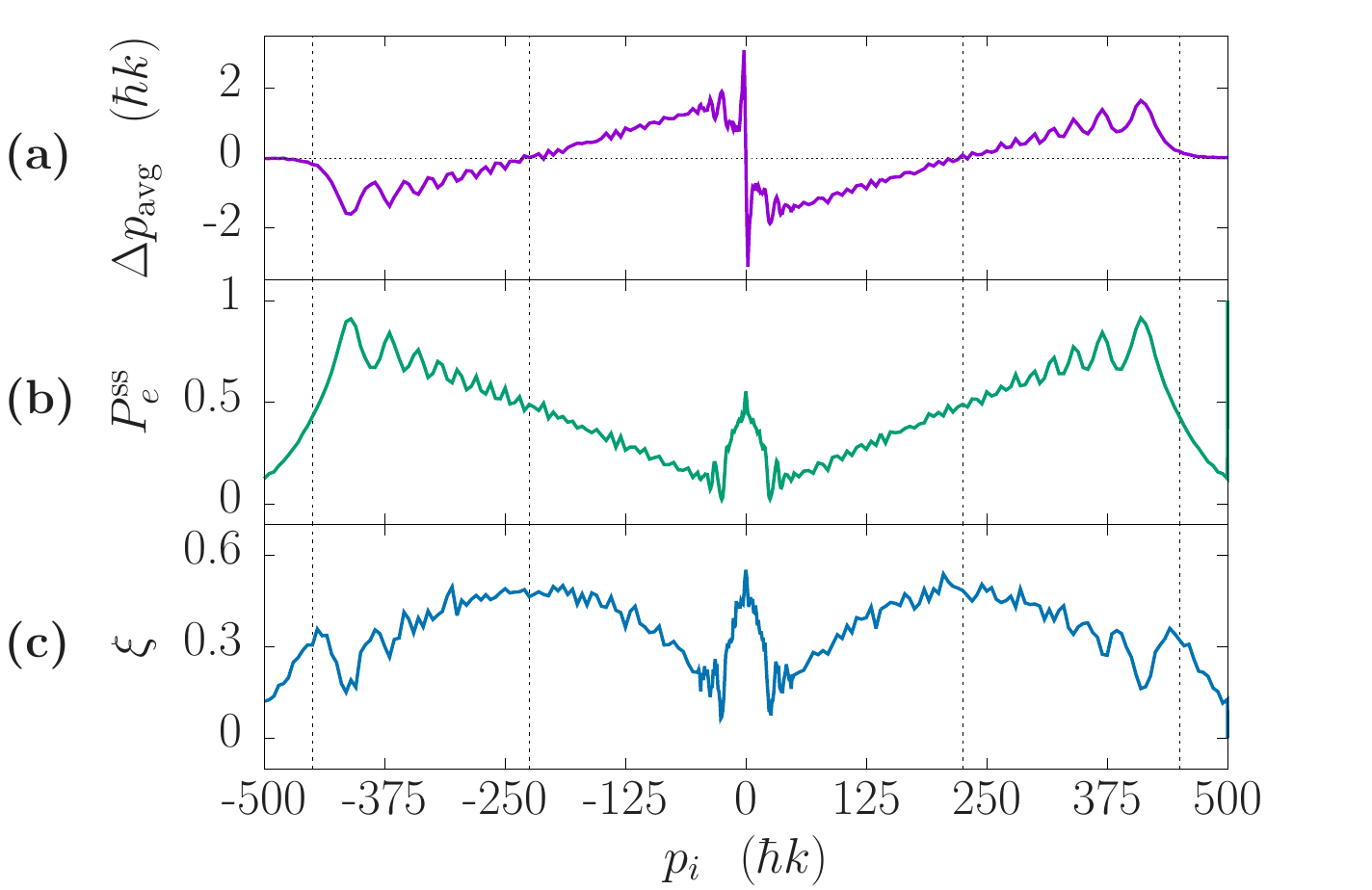}
  \caption{Various quantities for a particle that is subject to a single sweep with $\gamma \neq 0$ as a function of its initial momentum, $p_i$. (a) $\Delta p_\text{avg}$ (average impulse). (b) $P_e^\text{ss}$ (steady-state population). (c) $\xi$ (number of scattered photons). The vertical, dashed lines correspond to the four conditions $|kv| < \Delta_s/2$ (where $\Delta p_\text{rms}$ falls to zero) and $|kv| < \Delta_s/4$ (the capture range). Each point was averaged over 1000 trajectories. Values, in units of $\omega_r$, are: $\Delta_s = 1800$, $T_s = 1.0$, $\Omega_0 = 60, \gamma =1$.}
  \label{fig:diss}
\end{figure}

Figure \ref{fig:diss}(a) displays $\Delta p_\text{avg}$ for a large range of initial momentum states $p_i$. The parameters were chosen such that the time between the two laser interactions, $\tau_\text{res}$, obeyed $\tau_\text{res} \leq 1/\gamma$ for all $|k v_i| < \Delta_s/2$. We see that the overall effect of SWAP cooling yielded an impulse toward zero momentum for $|k v_i| < \Delta_s/4$ and an impulse away from zero momentum for $\Delta_s/4 < |k v_i|  < \Delta_s/2$. This motivates Eq.~(\ref{eq:captureRange}) as a characterization of the momentum capture range of SWAP cooling. For low momentum, the effects of Bragg oscillations, Dopplerons, and the ambiguous time-ordering of laser interactions results in momentum dynamics that differ from those in the high-velocity regime. 

Figures \ref{fig:diss}(b) and \ref{fig:diss}(c) present the steady-state excited state fraction, $P_e^\text{ss}$, and the average number of incoherent scattering events per sweep, which we call $\xi$, for the same parameters. We see that impulses with magnitudes of nearly $2 \hbar k$ are imparted for $|p_i|$ near $25 \hbar k$ with only $\sim 0.2$ scattering events per sweep. Moreover, the momentum states around $|p_i|=25\hbar k$ experienced an average force of $|\Delta p_\text{avg}/T_s| \approx 2 \hbar k \gamma$, which is roughly four times the cooling force that one expects from a radiation pressure force that fully saturates the atomic transition. This means that SWAP cooling can provide large cooling forces with a relatively low scattering rate. We investigate this useful feature in more detail in Section \ref{efficiency}.

\begin{figure}[!htb]
  \centering
  \includegraphics[width=1\linewidth]{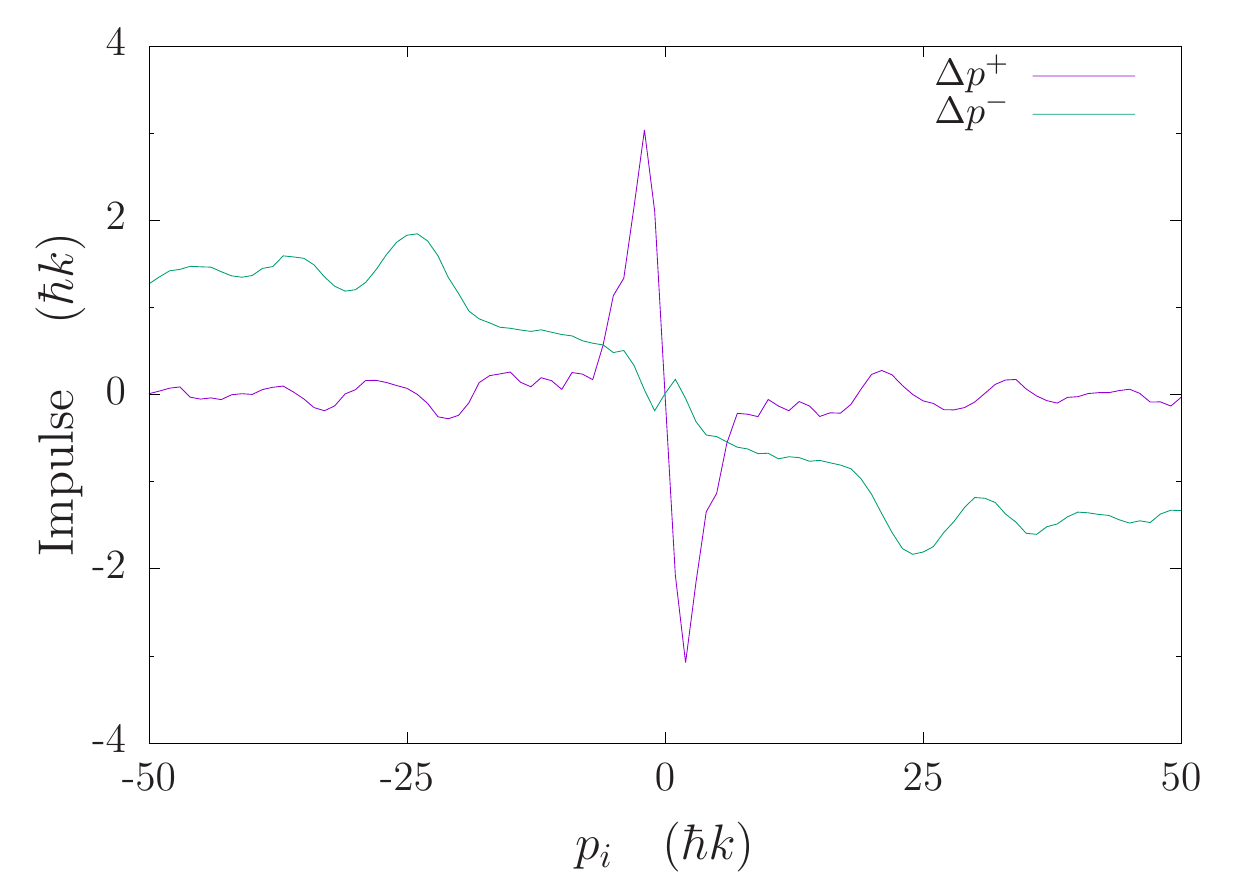}
  \caption{$\Delta p^\pm$ vs. $p_i$. The sharp, linear feature in $\Delta p^+$ is due to Bragg oscillations, which mix the particle between the $\ket{\pm p_i}$ states before it reaches resonance with the lasers. Each point is averaged over 1000 trajectories. Values, in units of $\omega_r$, are: $\Delta_s = 1800$, $T_s = 1.0$, $\Omega_0 = 60, \gamma =1$.}
  \label{fig:force_sum_and_difference}
\end{figure}

The effect of Bragg oscillations at low $p_i$, which manifests as a sharp, linear feature in Figure \ref{fig:diss}(a), is elucidated by considering the effect of switching the sweep direction. As described in Appendix \ref{bragg}, Bragg oscillations can mix the particle between the $\ket{\pm p_i}$ states before the particle resonates with the lasers, so the net $\Delta p_\text{avg}$ is independent of the sweep direction. To compare $\Delta p_\text{avg}$ with and without these oscillations, we define impulses $\Delta p^\pm$ in the following way:
\begin{equation}
\label{eq:sum_diff_force}
	\Delta p^\pm \equiv \frac{(\Delta p_\text{avg})_\text{pos} \pm 
    	(\Delta p_\text{avg})_\text{neg}}{2}
\end{equation}
in which the subscripts ``pos'' and ``neg'' refer to the sign of the ramp, {\em i.e.}, red to blue or blue to red detuning, respectively. By symmetry, we expect the effect of Bragg oscillations to be present in $\Delta p^+$ and to cancel in $\Delta p^-$. Figure \ref{fig:force_sum_and_difference} displays $\Delta p^\pm$ as a function of the initial momentum, $p_i$. The substantial effect of Bragg oscillations were seen to be present in $\Delta p^+$ for $|p_i| < 7 \hbar k$. For $p_i$ nearer to the high-velocity regime, which is $|p_i| > 30.5 \hbar k$ for this set of parameters, Bragg oscillations do not play a significant role, and we observe $\Delta p^+$ tending to zero, and $\Delta p^-$ tending to $\pm 1.5 \hbar k$.

\section{Temperature limit as $\gamma \rightarrow 0$} \label{QJ simulation results}
 \label{temperature_gamma_zero}

\begin{figure}[!htb]
  \includegraphics[width=\linewidth]{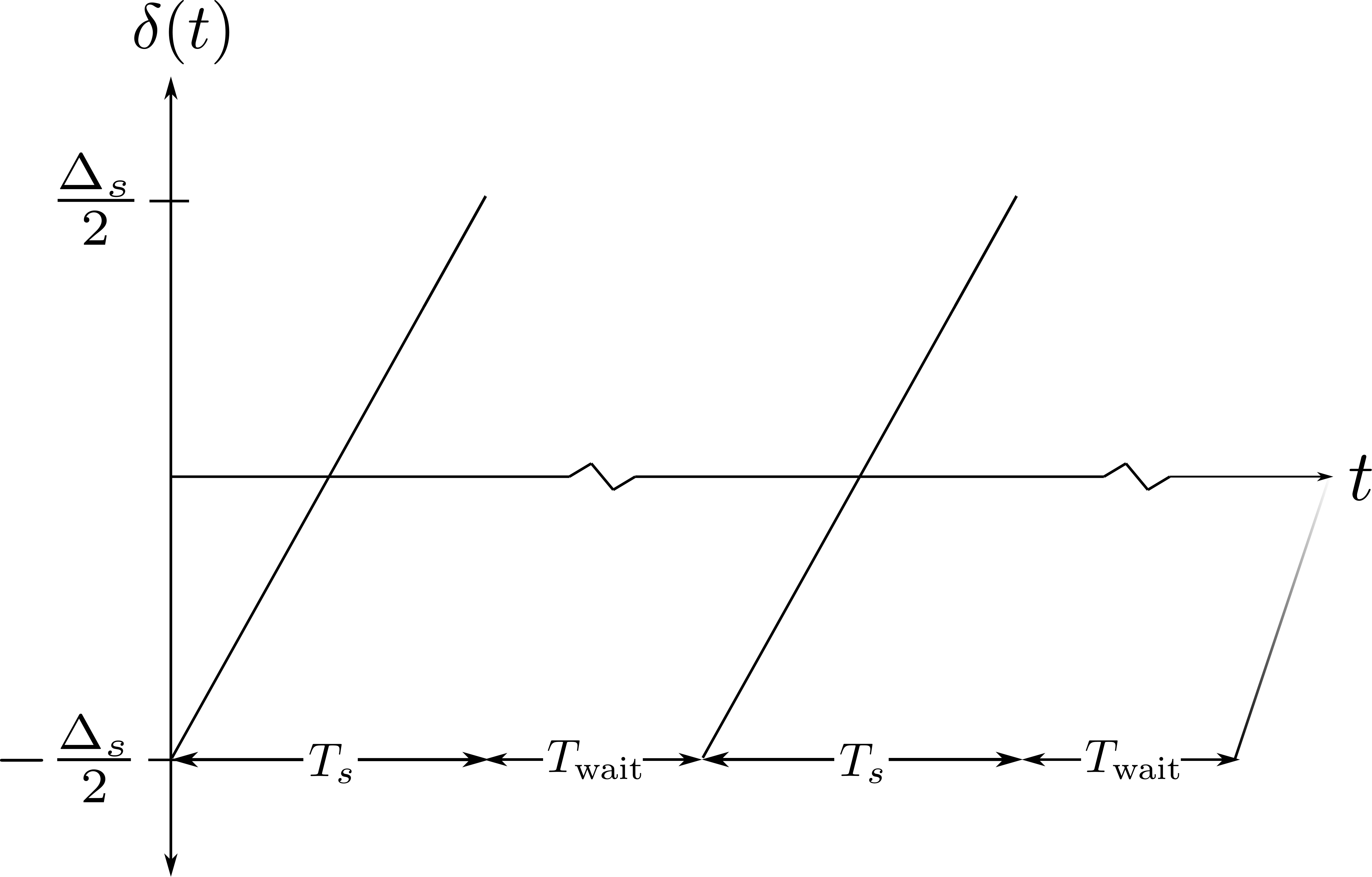}
  \caption{Frequency profile for the simplified ``sweep-wait'' cooling scheme.}
  \label{fig:simplified_cooling}
\end{figure}

\noindent We now provide the results of a SWAP cooling simulation that includes the effects of dissipation, but in the limit $\gamma \rightarrow 0$. To accomplish this, we modifed the frequency profile to be a series of single sawtooth ramps with period $T_s \ll 1/\gamma$, each separated by a time $T_\text{wait} \gg 1/\gamma$, as shown in Figure \ref{fig:simplified_cooling}. This ``sweep-wait'' scheme effectively mimics the cooling process for a particle with an ultranarrow linewidth. It also reduced the required computation time by allowing i) $\Delta_s$ to be small and ii) the use of the analytical expression for free-space spontaneous decay. We modeled the limit $T_\text{wait} \rightarrow \infty$ and $\gamma \rightarrow 0$ by setting $\gamma=0$ and projecting any remaining excited state population to the ground state, along with simulating any accompanying momentum recoil.

\begin{figure}[!htb]
  \includegraphics[width=\linewidth]{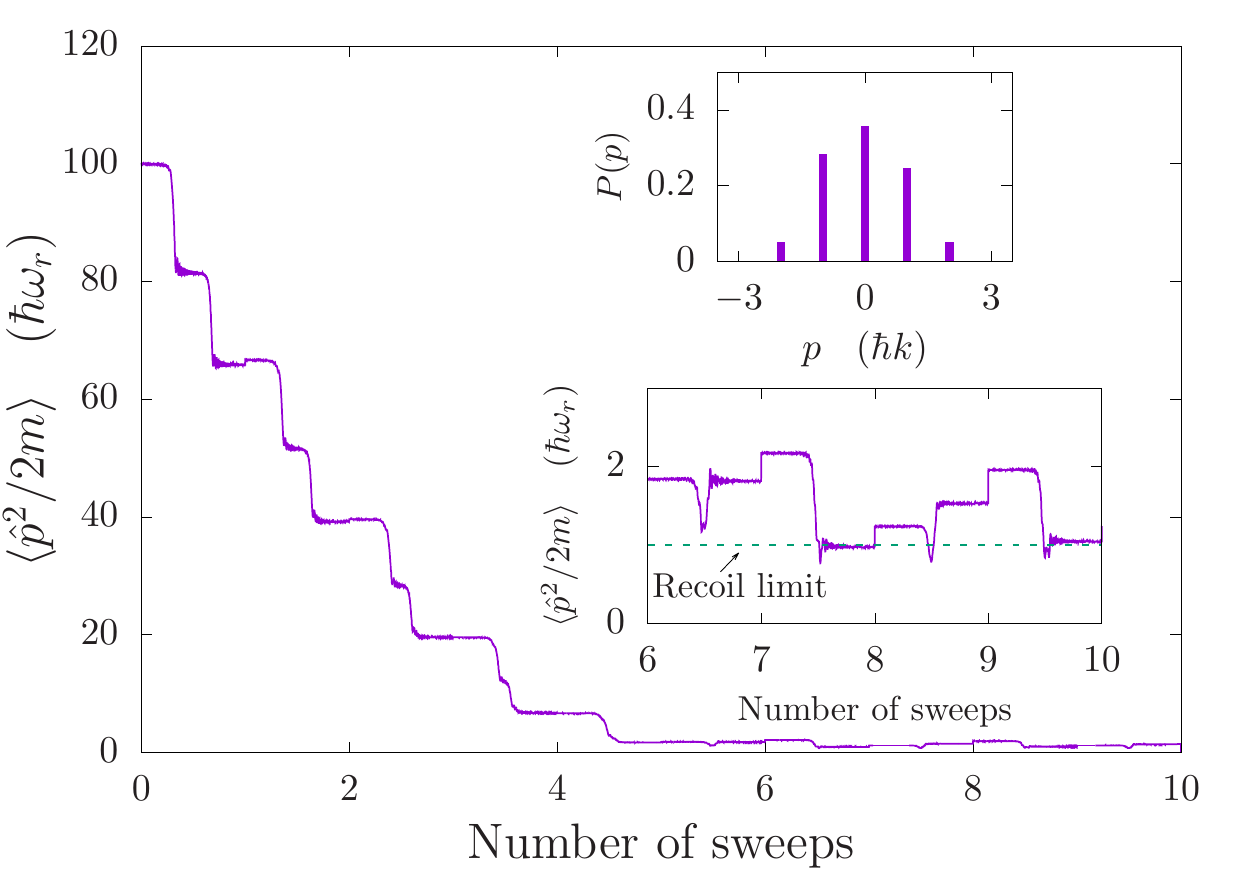}
  \caption{Cooling of a particle that begins in the state $\ket{g,10 \hbar k}$ to the recoil limit after 5 sweep-wait cycles. Top inset: A snapshot of the momentum distribution $P(p)$ halfway through the $7^\text{th}$ sweep. Bottom inset: A closer look at the cooling trajectory once the system reached equilibration. The recoil limit, $2T_r$, is included as a horizontal, dashed line. This curve is the average of 100 trajectories. Values, in units of $\omega_r$, are: $\Delta_s = 100, T_s = 60$, and  $\Omega_0 = 2$.}
  \label{fig:recoil_limit}
\end{figure}

Figure \ref{fig:recoil_limit} presents the evolution of a particle that started in the state $\ket{g,10 \hbar k}$ under this cooling scheme. At the end of this simulation, the temperature of the particle was observed to asymptote to a value near the recoil limit, $2 T_r$, where 
\begin{equation}
    	k_BT_r \equiv
    	\frac{(\hbar k)^2}{2m} =
        \hbar \omega_r,
\end{equation}
and $k_B$ is Boltzmann's constant. We use the variance in~$p$ as a measurement of the 1D temperature, $T$, {\em i.e.},
\begin{equation}
\label{eq:variance}
	\frac{\sigma_p^2}{2m} \approx 
    	\frac{\langle p^2 \rangle}{2m} = 
        \frac{1}{2} k_B T.
\end{equation}

\noindent It should be noted that not all steady state solutions are Gaussian in nature, but are centered on zero momentum. Thus, the first equality in Eq.~(\ref{eq:variance}) universally holds once the system has equilibrated.

\begin{figure}[!htb]
  \includegraphics[width=\linewidth]{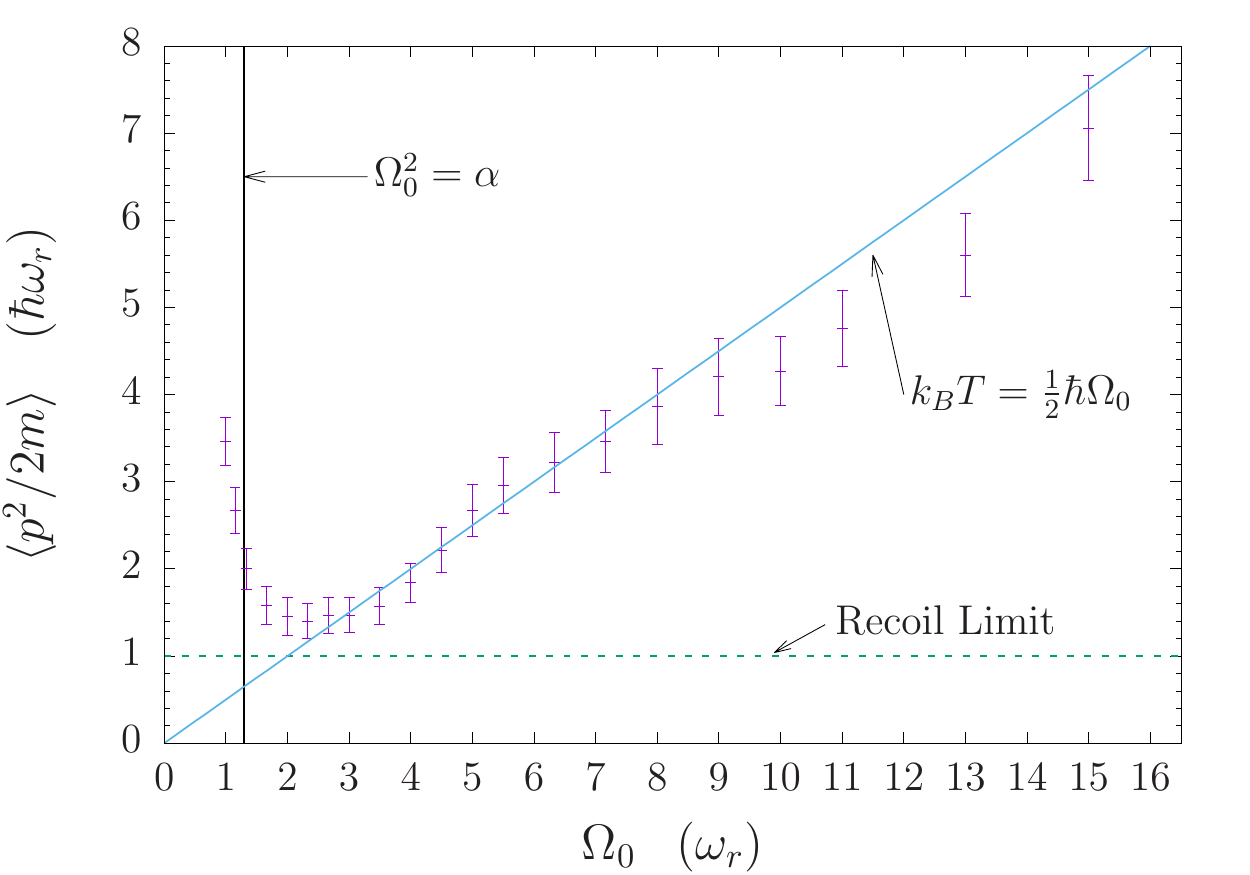}
  \caption{Stationary values of $\langle p^2/2m \rangle$ as a function of $\Omega_0$ in the sweep-wait sequence. The vertical, solid line labels the value of $\Omega_0$ that divides the diabatic and adiabatic regimes. The recoil limit is represented as a horizontal, dashed line. Deep within the adiabatic regime, the temperature scales linearly with $\Omega_0$. Values, in units of $\omega_r$, are: $\Delta_s = 100, T_s = 60.$ Each point is averaged over 500 trajectories.}
  \label{fig:omega_plot}
\end{figure}

It is also interesting to consider how the final temperature scales with $\Omega_0$. Figure \ref{fig:omega_plot} displays this relationship in both the diabatic and adiabatic regimes. A minimum temperature was found just within the adiabatic regime. From the numerics, the temperature was observed to follow a linear relationship
\begin{equation}
	k_B T = \frac{1}{2} \hbar \Omega_0
\end{equation}

\noindent in the adiabatic regime, which we attribute to decreased time ordering between adiabatic transfers and the particle spending more time in the excited state, hence more scattering events. This linear relationship held for at least twice the domain of Figure \ref{fig:omega_plot}. An increase in temperature was observed in the diabatic regime, which we attribute to both a reduction in conservative forces as described in Section \ref{conservative_forces}, and the presence of a significant excited state fraction at the end of each sweep, leading to more diffusion from spontaneous emission.

\section{Cooling efficiency}
\label{efficiency}

\noindent We now compare the efficiency of SWAP cooling with Doppler cooling. We define ``cooling efficiency'' here as the energy carried away from the system per scattering event. This choice is motivated in part by the potential application of SWAP cooling to systems where closed cycling transition may not be accessible and therefore a large number of spontaneous emission events are undesirable. We calculate the cooling efficiency by preparing an ensemble of particles in a specific initial state, $\ket{g,20\hbar k}$, and subsequently applying both Doppler and SWAP cooling for comparison. Here, we used the SWAP cooling scheme presented in Figure~\ref{fig:sweep}, in which dissipation is included at all times.

Clearly, many fewer scattered photons are required for SWAP cooling to reduce the energy and to bring the system close to equilibrium. In fact, in the simulation, SWAP cooling was able to remove up to an average of $5 \hbar k$ of momentum per scattered photon. As a consequence, in comparison to Doppler cooling's ideal cooling efficiency, we deduce that SWAP cooling promises to be well-suited to cooling particles where the adverse affects of spontaneous emission are significant, such as those that lack closed cycling transitions.

\begin{figure}[!htb]
  \centering
  \includegraphics[width=1.05\linewidth]{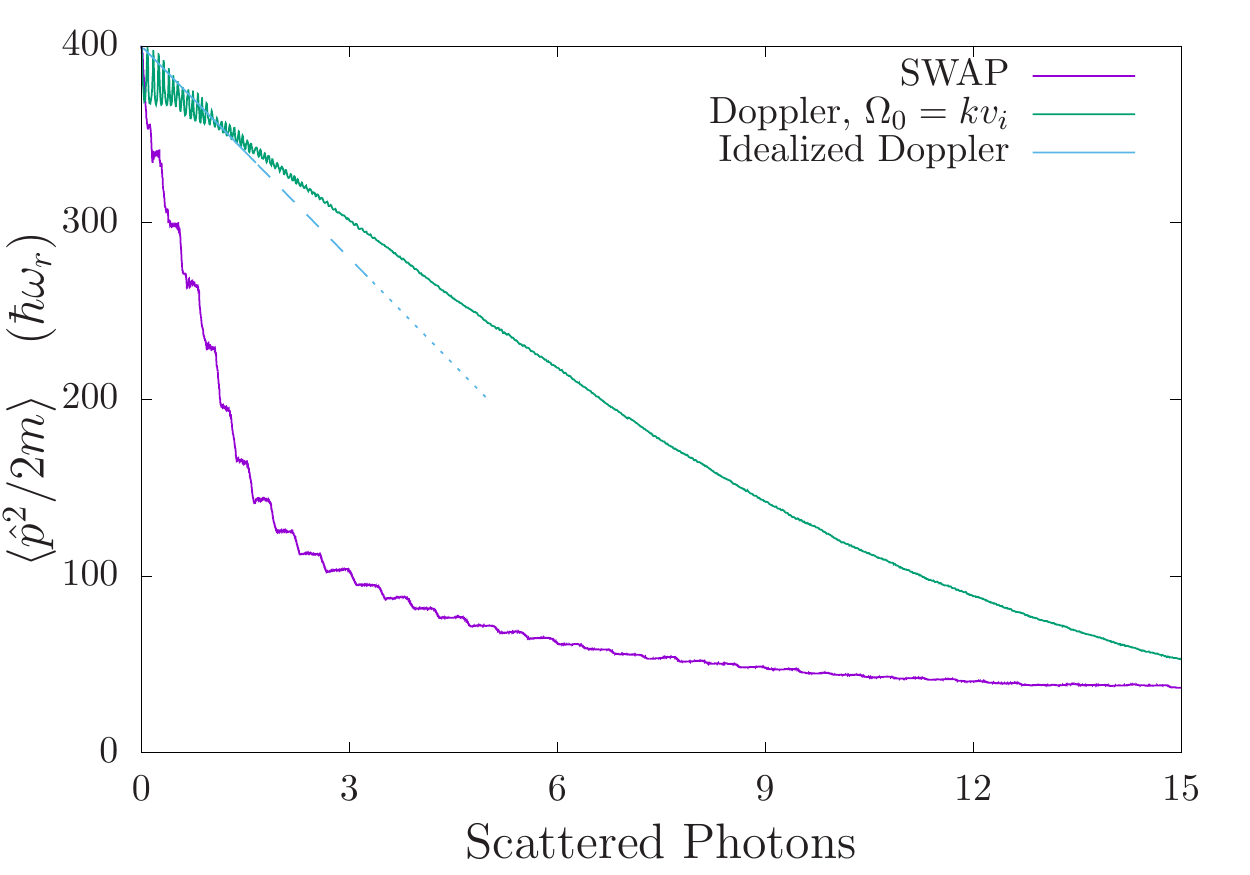}
  \caption{$\langle \hat{p}^2 \rangle / 2m$ vs. the average number of scattered photons for SWAP cooling and Doppler cooling. For both methods, $\omega_r=\gamma = 1$. The SWAP parameters, in units of $\omega_r$, are: $\Omega_0 = 28, \Delta_s = 391, T_s =1$, and Doppler parameters are: $\Omega=40, \delta=-40$. The SWAP cooling data is over 38 sweeps. Each curve is averaged over 1000 trajectories. We extract an average cooling efficiency for SWAP cooling of up to $5 \hbar k$ per scattered photon.}
  \label{fig:efficiency}
\end{figure}


\section{Conclusion}\label{conclusion}

\noindent The SWAP (sawtooth wave adiabatic passage) procedure proves to be a robust and simple cooling mechanism. Our analysis of both the coherent and dissipative dynamics of two-level particles suggests that it has numerous applications, such as to systems with narrow linewidth or no closed cycling transitions. We have shown its ability to cool particles to the recoil limit while simultaneously inducing larger cooling forces and maintaining a higher cooling efficiency than Doppler cooling.

In the future, it will be interesting to further elaborate on the concept of a cooling efficiency, {\em i.e.}, the removal of a system's energy and entropy per scattering event, in the general context of laser cooling theory. For example, the information associated with the momentum of a particle initially in the state $\ket{g,p_i}$ may be encoded in the time record of spontaneously emitted photons. This provides significant motivation for a  systematic analysis of entropy dynamics in the various implementations of laser cooling that may prove to be an insightful and useful endeavor for the laser cooling and atomic physics community.

The scope of the calculations presented in this paper were limited to two-level atoms moving along one-dimension. It will be interesting to consider more general atomic systems and more general geometries of laser and trapping fields. In particular, SWAP cooling may be applied to magneto-optical trapping \cite{swap_mot}, to optical lattices, and to general multilevel laser cooling strategies \cite{greve}. Furthermore, while it is important to recognize that while our calculations are fully quantum mechanical (consisting of a complete description of the internal and external variables) with the inherent advantages of not having to make approximations, it will also be useful to develop and validate semiclassical methods where many-sweep cycles can be treated with higher efficiency. This will be important to treat more massive systems such as complex molecules.

We would like to thank Athreya Shankar and John Cooper for many useful discussions. All authors acknowledge financial support from DARPA QuASAR, ARO, NSF PFC, and NIST. JRKC
acknowledges financial support from NSF GRFP. This work was supported by NSF PFC grant number PHY
1734006, DARPA Extreme Sensing, and NIST.

\bibliographystyle{apsrev4-1}
\bibliography{main}


\appendix

\section{Resonance times and intervals in the adiabatic limit}
\label{tres}

\noindent In the adiabatic limit ($\Omega_0^2 \gg \alpha$), most of the population transfer for a motionless particle occurs around $t=0$ for a detuning profile $\delta(t) = \alpha t$, as this is when the laser is in resonance with the particle. By energy conservation (see Eq.~(\ref{eq:doppleron_timing}) with $n=0$), motional degrees of freedom translate this ``resonance period'' for a particle in the state $\ket{g,p}$ interacting with a right-traveling wave to the time that satisfies
\begin{equation} \label{eq:righttime}
    \alpha t_\text{right} - \omega_r(2\beta+1) = \alpha t_\text{right} - k v - \omega_r  = 0,
\end{equation}

\noindent where $\beta \equiv p/\hbar k$. For high-velocity particles ($kv \gg \omega_r$), the resonance period agrees with the intuitive Doppler shift result. However, the additional recoil term is important for low velocity ($kv \simeq \omega_r$) particles. A similar result can be found for the interaction of a ground-state particle with a left traveling wave ($\beta \rightarrow -\beta$), which translates the resonance period to
\begin{equation} \label{eq:lefttime}
    \alpha t_\text{left} - \omega_r(-2\beta +1) = \alpha t_\text{left} + k v - \omega_r  = 0.
\end{equation}

\noindent The resonance periods for a particle starting in the excited state can be found with the substitution $\omega_r \rightarrow - \omega_r$ on the additional recoil term. It is easily shown that adiabatic transfer with motion has the same probability as the motionless case \cite{zener}.

In the high-velocity regime (see Eq.~(\ref{eq:highvelocityregime})), there is one stimulated absorption and one stimulated emission per sweep in SWAP cooling. From Eqs.~(\ref{eq:righttime}) and~(\ref{eq:lefttime}), the time separating these resonant phenomena for a particle with initial velocity $v_i$, labeled $\tau_\text{res}$, is
\begin{equation}
\label{eq:tres}
	\tau_\text{res} = \frac{2(kv_i - 2\omega_r)}{\alpha}.
\end{equation}

\noindent It should be noted that (\ref{eq:tres}) is only valid in the adiabatic limit, since the time associated with the stimulated absorption recoil has been included. An understanding of these resonance times and intervals provides insight into the particle dynamics for various momentum states in SWAP cooling.


\section{Doppleron Regime} \label{doppleron} 

\begin{figure}[!htb]
  \begin{center}
  \includegraphics[width=1\linewidth]{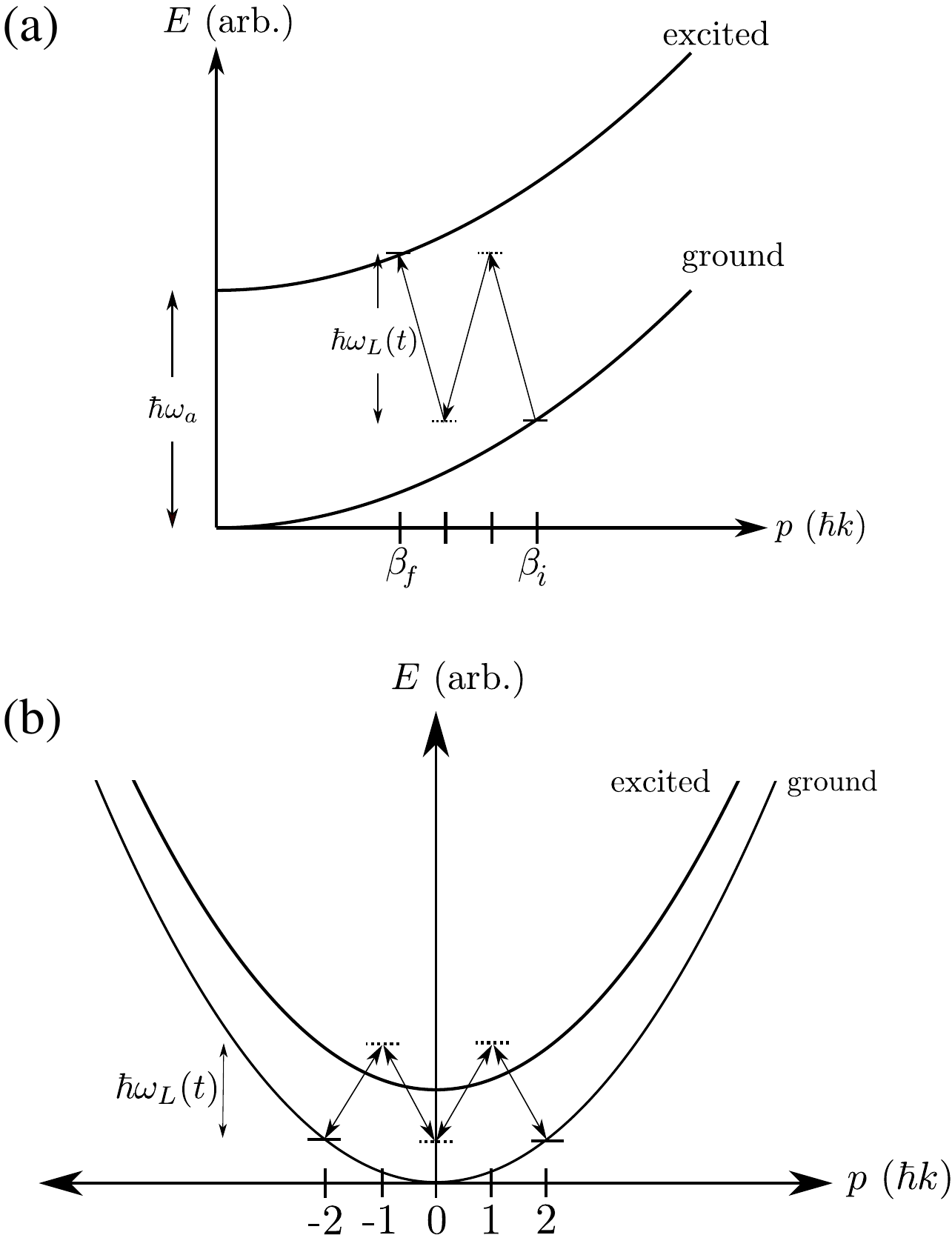}
  \end{center}
\caption{(a) A $1^\text{st}$-order Doppleron resonance, characterized by the condition $\beta_f = \beta_i - 3$. (b) A Bragg resonance between the states $\ket{g, \pm 2\hbar k}$.}
  \label{fig:multip}
\end{figure}

\noindent As the laser frequencies are swept, multi-photon transitions often called Dopplerons come to resonance even when single-photon transitions do not.  As shown in Figure \ref{fig:multip}(a), an $n^\text{th}$-order Doppleron process is characterized by the absorption of $n+1$ photons from one beam and the emission of $n$ photons into the other, resulting in a $(2n+1) \hbar k$ net momentum transfer to the particle and the particle being left in the opposite internal state from which it started \cite{minogin_dopplerons}. In general, Doppleron resonances can occur both before and after the single-photon resonances, and their existence has the potential to substantially affect particle dynamics outside the high-velocity regime.

Doppleron resonances only occur for specific values of laser detuning. For a particle with initial momentum $p_i= \beta_i \hbar k$, the time of an $n^\text{th}$-order Doppleron resonance can be found from energy conservation:
\begin{equation}
\label{eq:doppleron_energy}
	 \beta_i^2 \hbar \omega_r + (n+1) \hbar \omega_L(t) = n \hbar \omega_L(t) + 	   \hbar \omega_\text{a} + \beta_f^2 \hbar \omega_r.
\end{equation}

\noindent We have neglected the AC Stark shift here for simplicity. In Eq.~(\ref{eq:doppleron_energy}), the terms proportional to $n$ are the energies of the photons being absorbed and emitted, $\hbar \omega_\text{a}$ is the transition energy, and the terms proportional to $\omega_r$ are the initial and final kinetic energies of the particle. 

Paired with the momentum condition $\beta_f = \beta_i - (2n+1)$, where $\beta_f$ is the particle's momentum after the Doppleron process, the time, $t_n$, of an $n^\text{th}$-order Doppleron resonance for a lab-frame laser detuning profile $\delta_n(t) =\alpha t_n$ obeys
\begin{align}
	\label{eq:doppleron_timing}
	\alpha t_n & = 
    	-(2n+1)kv_i + (2 n+1)^2 \omega_r \notag \\
    & = -\left(n+\frac{1}{2}\right)(kv_i + k v_f).
\end{align}

Regardless of the Rabi frequency, multiphoton effects become relevant for low momentum states. From numerical methods, the 1$^\text{st}$-order splitting in a dressed state picture (see Figure \ref{fig:4eval}) behaves as

\begin{equation}
    \Omega^{(1)} \approx \frac{\Omega_0^3}{16(kv_i- 3 \omega_r)^2},
\end{equation}

\noindent which suggests that the probability of adiabatically passing through a first-order Doppleron resonance is \cite{zener}
\begin{equation}
    P_a^{(1)} = 
    1-\exp
    \left[
    - \frac{\pi}{512} \frac{\Omega_0^6}{\alpha (kv_i - 3 \omega_r)^4}
    \right].
\end{equation}

\noindent The size of the argument in this exponential can be used to define the conditions for which Dopplerons are relevant for a particle with velocity $v_i$. Because SWAP cooling requires Eq.~(\ref{eq:atrans}) to be satisfied, we choose to define the Doppleron regime as follows:
\begin{align}
    |\Omega_0| &> |kv_i - 3 \omega_r|	\qquad \qquad (\text{Doppleron regime}). \label{eq:regimes}
\end{align}

\noindent Comparing this result with Eq.~(\ref{eq:highvelocityregime}), we see that the Doppleron regime describes nearly all remaining states outside the high-velocity regime. Of course, there exist similar conditions for higher-order Doppleron resonances.

Increasing $\Omega_0$, {\em i.e.}, allowing for more Doppleron resonances, may decrease the amount of time it takes to reach a steady-state temperature, as they allow for more momentum transfer. However, the final temperature increases with $\Omega_0$ (see Figure \ref{fig:omega_plot}). Thus, minimum temperatures with a high capture range may be achieved by dynamically changing the Rabi frequency as the particle is cooled over time.


\section{Bragg Oscillations}
\label{bragg}

\noindent Momentum states experience nontrivial effects due to Bragg oscillations, i.e., particle scattering from a light grating (the standing wave). These transitions, which are between states with $\pm \beta \hbar k$ momenta where $\beta$ is an integer, occur at a rate \cite{bragg}
\begin{equation}
\label{eq:bragg}
	\Omega_{B,\beta}(t) \approx
    	\frac{|\Omega_0|^{2 \beta}}{4^\beta(8 \omega_r)^{\beta-1}[(\beta-1)!]^2\delta(t)^\beta},
\end{equation}

\noindent provided that $|\delta(t)| > |\Omega_0| > \gamma.$  Figure \ref{fig:multip}(b) displays the energy diagram for a Bragg resonance between the $\ket{g, \pm 2 \hbar k}$ states. If there are many oscillations, it is likely that system noise will cause the oscillations for different particles to become out of phase, leaving, on average, half of the atoms in either momentum state. This phenomenon is an important consideration in the analysis of the impulses in Figure \ref{fig:force_sum_and_difference}.

Integrating Eq.~(\ref{eq:bragg}) and dividing by $2 \pi$ gives the total number of $\beta^\text{th}$-order Bragg oscillations between times $t_i$ and $t_f$:
\begin{align}
\label{eq:bragg_osc}
	N_\beta(t_i,t_f) & =
    	\frac{1}{2\pi} \left| \int_{t_i}^{t_f}
        	\Omega_{B,\beta}(t') \, dt' \right| \\ 
    & \propto \left| \int_{t_i}^{t_f} \frac{1}{\delta(t')^\beta} \, dt'. \right|\notag
\end{align}

\noindent The quantities $t_i$ and $t_f$ are related to the times of stimulated absorption/emission during the sweep process, as these will carry the particle away from the states that are experiencing Bragg oscillations.

To provide a few interesting examples, $1^\text{st}$-order Bragg transitions affect the dynamics of the initial states $\ket{\psi}_0 = \ket{g, \hbar k}, \ket{g,3 \hbar k}$. From Eq.~(\ref{eq:bragg_osc}) and using Eqns.~(\ref{eq:righttime}) and~(\ref{eq:lefttime}), 
\begin{align}
	N_1 & \approx \frac{\kappa}{8 \pi} \ln
    \left| 
    \frac{2 \omega_r}{\Delta_s}
    \right|, \quad \ket{\psi}_0 = \ket{g,\hbar k}; \\
    N_1 & \approx \frac{\kappa}{8 \pi} \ln
    \left| 
    \frac{\Delta_s}{6 \omega_r}
    \right|, \quad \ket{\psi}_0 = \ket{g,3 \hbar k}. \notag
\end{align}

\noindent The Bragg oscillations occur before the SWAP resonances for $\ket{g,\hbar k}$ and after the SWAP resonances for $(\ket{g,3 \hbar k})$. Although $N_1$ scales logarithmically with $\Delta_s$ in both cases, a sufficiently large sweep range will promote multiple Bragg oscillations.

\end{document}